# Variable-Based Network Analysis of Datasets on Data Exchange Platforms

**Teruaki Hayashi, Yukio Ohsawa**
Department of Systems Innovation, School of Engineering
The University of Tokyo, Japan
hayashi@sys.t.u-tokyo.ac.jp, ohsawa@sys.t.u-tokyo.ac.jp

## Abstract

Recently, data exchange platforms have emerged in the digital economy to enable better resource allocation in a data-driven society, which requires cross-organizational data collaborations. Understanding the characteristics of the data on these platforms is important for their application; however, the structures of such platforms have not been extensively investigated. In this study, we apply a network approach with a novel variable-based structural analysis to the metadata of datasets on two data platform services. It was noted that the structures of the data networks are locally dense and highly assortative, similar to human-related networks. Even though the data on these platforms are designed and collected differently, depending on the use objectives, the variables of heterogeneous data exhibit a power distribution, and the data networks exhibit multi-scaling behavior. Furthermore, we found that the data collection strategies of the platforms are related to the variety of variables, density of the networks, and their robustness from the viewpoint of sustainability and social acceptability of the data platforms.

## 1 Introduction

In recent years, data have been used as an exchangeable economic resource, and new businesses have emerged because of data trading across fields [Balazinska et al., 2011; Liang et al., 2018]. However, researches and businesses require a large amount of data, and the lack of data has become a critical and urgent issue to be solved. The use of data exchange platforms can help ensure that more data are available, even if at a cost, and combinable. Consequently, various forms of data exchange platform services (hereafter, data platforms) have been launched worldwide [Ohsawa et al., 2013; Mano, 2016; Zheng et al., 2017; Dai et al., 2019]. In such services, a marketplace on the Internet is developed in which companies (buyers) who want to buy data in different areas and companies (suppliers) who want to sell data can freely participate. Data platforms are thus new types of marketplaces that have recently emerged in the digital economy, and their structures have not been investigated extensively until now. In these platforms, various types of data are integrated, which compete among one another, and an appropriate technique to analyze such an ecosystem has not been established. The statistics and dynamics of those data platforms must be understood to solve complex problems that require cross-organizational data collaborations. In the existing studies, the types of relationships that appear among different data and the mechanisms behind the data and the platforms have not been clarified. Previous researchers have investigated and discussed that no typical strategy or common knowledge of data values exists [Stahl et al., 2014; Spiekermann, 2019]. Understanding the characteristics of data on such platforms is important to popularize the use of data platforms and enable better resource allocation in a data-driven society. Therefore, in this study, we pose and answer the following question: *What are the topological characteristics of data platforms?*

To understand the aspects and interactions of data on the platforms, it is first necessary to understand the structure of the data and the associated relationships. In this study, we assumed the data handled on the platforms as the population and the collected metadata as samples. On this basis, we quantitatively investigated the structural characteristics and relationships of the data on data platforms. The key contributions of this study are as follows: The characteristics of the macro-level data structures and networks from the empirical data of two different platforms are analyzed. Moreover, the robustness of the data platforms is evaluated from the viewpoint of sustainability and social acceptability. Because the data platforms are emerging marketplaces, it is likely that an ecosystem may not function suitably when some data are lost or data users/holders leave the platforms. Consequently, understanding the behaviors and interactions of data on a platform could elucidate the combinability of the data and provide valuable insight to the users/providers of data platforms.

## 2 Proposed Approach

### 2.1 Motivation and Related Studies

The main purpose of this study was to understand the structural characteristics of the data on data exchange platforms and discuss the mechanisms behind such structures. In particular, we attempted to interpret the macro-level law of data, which is the common rule behind the data when a population of data is observed. Moreover, to evaluate the sustainability of data platforms in the society, we considered the robustness of the data platforms. To this end, two methods were used to understand the characteristics of the data platforms: analysis of the structures of the data by focusing on the variables, and determining the network characteristics of the data by considering the variables.

First, the variables in the data were used to understand the structure of the data across fields. A variable is a logical set of attributes, and an attribute is a characteristic of a person or a thing [Babbie, 2016]. For example, the variable "weather" has attributes of "sunny," "cloudy," or "rainy," and the variable "gender" has attributes of "women" and "men." We described the universal features of the data by using variables and discussed the structural differences/similarities of the data. Second, a network analysis was performed. To reduce the biases of mismeasured variables, different data sources can be combined to answer research/business questions instead of relying on a single data source [Ridder and Moffitt, 2007]. Although some technical difficulties are encountered in linking heterogeneous data such as schemas [Erhard and Do, 2000; Pang et al., 2015], the use of variables has been demonstrated as being effective for combining data. From the perspective of data exchange on the platforms, the data can be considered as nodes, and the variables can be considered as the links between the different datasets. By assuming that the network consists of the interactions of the data, the structural characteristics and their relationships can be elucidated.

The existing studies on data platforms and data marketplaces employed the game theory [Sooksatra et al., 2018], business roles and marketability [Stahl et al., 2014; Quix et al., 2017], market models [Ohsawa et al., 2013], trading and pricing models [Shen et al., 2016; Cao et al., 2017], blockchain based data trading systems [Dai et al., 2019], data protection and digital rights [Tassel, 2006; Liang et al., 2018], and privacy issues [Niu et al., 2017]. However, the structural characteristics of the data platforms have not been considered by using a network approach focusing on the variables. Although some researchers applied the network approach to linked data [Guéret et al., 2012], their motivation was to evaluate the link quality and not to understand the structures of the data networks. Furthermore, although Ruijer et al. [2017] discussed the platforms of open data, they did not focus on the variables in the data and performed case studies instead of a network analysis.

## 2.2 Datasets

To address our research questions, we examined the metadata of datajacket.org[1] and D-Ocean[2]. Datajacket.org collects the metadata in Data Jacket (DJ)—a framework for describing the summary information of data while maintaining the confidentiality of the data [Ohsawa et al., 2013]. The summary information of the data pertains to the explanatory text regarding the data, for instance, variable names, format, data and sharing conditions (12 description items). The use of DJ helps understand the types of data that exist in different domains and the types of information that the data include, even if the contents of the data cannot be made public. We obtained 1316 DJs as metadata that were widely collected from business persons, researchers, and data scientists. Description items of the variables (variable labels) were extracted from the DJs. A variable label is explanatory text regarding the variables specific to the data, e.g., "latitude," "longitude," "name," or "age," which are the most representative attributes of data. D-Ocean—the social networking service of the data marketplace—provides metadata in their own format. We extracted the description items of the "column definition" from the metadata of D-Ocean, which correspond to the variables. A total of 3749 metadata having variables was extracted from D-Ocean. Although we have access to the complete metadata of datajacket.org and D-Ocean, the dataset of D-Ocean is larger, and a bias thus exists in the number of samples. To evaluate the difference between the two platforms equally, we conducted random sampling for the D-Ocean dataset, and acquired a dataset with nearly the same amount of data (831 metadata) as that of datajacket.org. Table 1 summarizes the two datasets described above.

|  | datajacket.org | D-Ocean |
|---|---|---|
| Total # of data | 1316 | 3749 |
| Total # of variables | 9022 | 73214 |
| # of types of variables | 6594 | 2678 |
| Max. # of variables in data | 119 | 235 |
| Min. # of variables in data | 1 | 1 |

Table 1: Data obtained from the data platforms

Using the data of these two platforms has the following two advantages. First, the datasets can be combined by arranging the variables included in the datasets. Although extensive metadata and many databases exist in specific fields, virtually no databases of metadata have been accumulated in accordance with the uniform description formats across different domains. For example, the metadata provided by data.gov[3] or the Humanitarian Data Exchange (HDX)[4] include data from various fields. However, these data do not have the descriptive items corresponding to the variables to allow the assessment of the data connections. The metadata of HDX have tags, which are provided as clues for searching, and they include the partial characteristics of the variables. However, tags are not exhaustive and not sufficient to consider the data combinability. Second, although the two considered datasets aim to enable the data exchange among different users, they have different strategies to collect data. Datajacket.org covers not only open data but also data owned by companies and individuals. In contrast, D-Ocean provides a platform for the cross-domain use of open data published by various institutions. In this study, we attempted to compare the differences in the structural characteristics by considering the data collection strategies.

---

[1] https://datajacket.org/?lang=english
[2] https://www.docean.io/en/index.html
[3] https://www.data.gov/
[4] https://data.humdata.org/

## 2.3 Analysis Methods and Metrics

First, we analyzed the structural characteristics of the data platforms by considering the distributions of their variables. We defined the probability of a variable appearing $m$ times as $p(m)$. However, $p(m)$ is small when $m$ is large, and very few $m$ exist for which $p(m) > 0$. Consequently, the double logarithmic graph of the frequency distribution is weak against the noise. Therefore, we used a rank-frequency plot, which is equivalent to the complementary cumulative distribution function, that is, $\wp(m) \equiv \int_k^\infty p(m')\, dm' \propto m^{-(\gamma-1)}$, where $\gamma$ is a power-law index [Newman, 2007]. Every $\gamma$ is calculated considering the part of the distributions whose coefficient of determination is $R^2 \geq 0.97$. The subscripts {dj, do} represent the datasets from datajacket.org and D-Ocean, respectively.

Second, considering that a data platform is the place where people meet through data, we assumed that the interactions of the data on the platforms act as the network structures. In fact, several models using network approaches have been proposed for understanding social networking services [Palla et al., 2005]. Based on this assumption, we assumed the data platforms to be the data networks for understanding their topological characteristics. The combinability of the data depends on whether the data involved have common variables. To express this model, the data network that mediates the variables was considered as an undirected graph $G := (V, E)$. Here, $V$ is the set of nodes composed of data ($V = \{v_i | v_i \in \mathcal{D}\}$), $E$ is the set of edges ($E = \{e_{ij}\}$), and $\mathcal{D}$ represents the set of data ($v_i \in \mathcal{D}$). An edge is established when the same variables appear simultaneously in a pair of data and is represented as $e_{ij} = \{v_i, v_j\}$ iff $v_i, v_j \in \mathcal{D}, v_i \neq v_j$.

We used six features to evaluate the structure of the data networks: the average degree, density, average clustering coefficient, assortativity, average path length, and diameter. When the number of nodes ($|V|$) and number of links ($|E|$) are given, the average degree $\langle k \rangle = 2|E|/|V|$, and density $\rho = \langle k \rangle / (|V| - 1)$. The density is the ratio of the number of links to the number of all possible links with nodes in the network. The cluster coefficient describes the proportion of the links that are present between the neighboring nodes of a certain node. Assuming that $k_i$ is the number of links possessed by the $i$th node, and $L_i$ is the number of links present between the neighboring nodes of the $i$th node, the average cluster coefficient ($\langle C \rangle$) can be expressed as shown in (1).

$$\langle C \rangle = \frac{1}{|V|} \sum_{i=1}^{|V|} \frac{2L_i}{k_i(k_i - 1)} \quad (1)$$

The assortativity ($r$) indicates the degree of correlation between two neighboring nodes ($-1 \leq r \leq 1$). The degree correlation ($k_{nn}$) is a mapping between a node degree $k$ and the mean degree nearest neighbors of the nodes of degree $k$. The assortativity is the Pearson correlation coefficient of the degrees and $k_{nn}$. Assuming that $k_p$ and $k_q$ respectively indicate the degrees of the nodes $p$ and $q$ combined by link $s$, the assortativity can be expressed as in (2) [Newman, 2002]. If $r > 0$, the high-degree nodes tend to connect with high-degree nodes. In contrast, if $r < 0$, the high-degree nodes tend to connect with low-degree nodes. The assortativity values of biological system networks such as proteins or the food chain and engineering system networks such as the Internet tend to be negative. On the other hands, the assortativity values of networks that indicate human relationships, such as the co-author relationships for articles or performance relationships for films, tend to be positive [Newman and Park, 2003; Redner, 2008].

$$r = \frac{|E|^{-1} \sum_s k_p k_q - \left[|E|^{-1} \sum_s \frac{1}{2}(k_p + k_q)\right]^2}{|E|^{-1} \sum_s \frac{1}{2}(k_p^2 + k_q^2) - \left[|E|^{-1} \sum_s \frac{1}{2}(k_p + k_q)\right]^2} \quad (2)$$

The average path length is the mean distances between all the possible nodes and can be calculated using (3), where $d(v_i, v_j)$ denotes the shortest distance between $v_i$ and $v_j$. The diameter ($d^{\max}$) is the largest value of $d(v_i, v_j)$.

$$\langle d \rangle = \frac{1}{|V|(|V| - 1)} \sum_{i \neq j} d(v_i, v_j) \quad (3)$$

To evaluate the robustness of the data platforms, we removed the nodes from each network by a certain rate ($f$) and calculated the proportions of data connected to the largest component. $P_\infty(f)$ is the probability that when some nodes are removed at the rate of $f$, the other nodes belong to the largest component. Therefore, the size of the largest component is represented by $P_\infty(f)/P_\infty(0)$. In other words, the size of the largest component is maximum when $f = 0$ and minimum when $f = 1$. The loss of data was considered as the node removal, and on this basis, we discussed the differences between the platforms and network robustness pertaining to the data collection strategies. In the experiment, we compared the robustness of the random removal and the removal from high-degree nodes.

## 3 Analysis Results

### 3.1 Characteristics of Data and Variables

The variable-based data structures exhibit different and similar characteristics depending on the platforms. D-Ocean includes more data and variables; however, datajacket.org has 6594 types of variables, and D-Ocean has 2688 types (Table 1). In datajacket.org, 87.5% of the total variables appear once, and 8.0% appear twice. In other words, the variables that appear once or twice account for 95.6% of the total. In contrast, in D-Ocean, 54.2% of the variables appear once, and variables that appear once or twice account for 60.2%. The distributions of the variables of both datajacket.org and D-Ocean are in accordance with $p(m) \propto m^{-\gamma}$ (Fig. 1(a)), which becomes a power distribution ($\gamma_{\rm dj} = 2.30; \gamma_{\rm do} = 1.83$). In terms of the variables, $\gamma$ represents the degree of concentration on variables with low frequencies of appearance. The results show that the data on both platforms do not consist of many variables with high frequencies of appearance; instead, the data mainly consist of assorted and numerous variables with low frequencies of appearance. In addition, datajacket.org has a higher variety of variables than D-Ocean.

Next, we compared the proportions of the number of variables in each dataset. Approximately 94% of the total data is accounted for by the data with 1–15 variables in

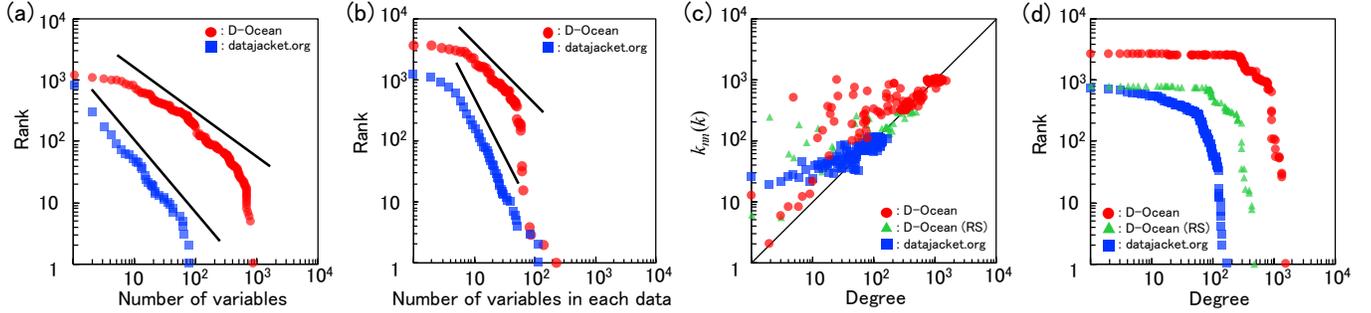

Figure 1: Topological characteristics of data platforms by metrics

datajacket.org; in D-Ocean, the data with 1–52 variables makes up 94% of the total data. Figure 1(b) shows the rank-frequency plot of the distributions of the number of variables in each data, designating the proportion of data with $l$ variables as $p(l)$. The numbers of holding variables are partially in accordance with $p(l) \propto l^{-\gamma}$ and exhibit a power distribution in both the platforms ($\gamma_{dj} = 3.00$; $\gamma_{do} = 1.97$), which means there is no average number of holding variables.

### 3.2 Topological Characteristics of Data Networks

Table 2 summarizes the largest components of the data networks. This study was focused on the largest components, because the other 441 components in datajacket.org were composed of ≤ 14 data. In contrast, although the D-Ocean network has 20 components, the second largest component including 865 nodes is a complete graph, and it does not have structural characteristics that can be discussed. Note that RS corresponds to the 831 datasets randomly sampled from D-Ocean, also represented by the subscript "rs."

The difference between the platforms occurs because the data pairs of D-Ocean have many variables in common. In terms of the average degrees, $\langle k \rangle_{do}$ is 13.8 times higher, and $\langle k \rangle_{rs}$ is 4.3 times higher than $\langle k \rangle_{dj}$. $\rho$ represents the global density, and $\langle C \rangle$ indicates the local density. $\rho_{dj}$ is low at 0.051, $\langle C \rangle_{dj}$ is high at 0.702, and $G_{dj}$ is locally dense and globally sparse. In contrast, $\rho_{do}$ is relatively high at 0.205, and $\langle C \rangle_{do}$ is high at 0.891, which represents the locally and globally dense characteristics of $G_{do}$. This phenomenon occurs because the pairs of data from datajacket.org have few variables in common, they have a low likelihood of connecting, and constitute a sparse network. In contrast, the data from D-Ocean share more variables in common and constitute a dense network. In addition, $r_{dj} = 0.489$ and $r_{do} = 0.589$, which expresses a high assortativity in both the networks. These results indicate that the networks consist of high-degree data connected to other high-degree data, and the low-degree data are connected to low-degree data. Figure. 1 (c) shows the distribution of the degree correlation ($k_{nn}(k)$). As discussed in the assortativity case, high positive correlations exist in both the networks. These characteristics suggest that both $G_{dj}$ and $G_{do}$ resemble the networks of human relationships more than the natural world or engineering system networks [Watts and Strogatz, 1998; Newman and Park, 2003; Newman, 2004]. In particular, $G_{dj}$ has a moderate positive correlation compared to $G_{do}$, and the distribution is scattered in higher-degree regions. In contrast, although $G_{do}$ has a large variance in the high-degree region, some nodes show a nearly perfect degree correlation from the lower- to higher-degree regions. The results suggest that not only are the hub nodes strongly connected each other, each node is connected to the data having nearly the same degree in $G_{do}$. This result is strongly influenced by the data collection strategies, as discussed in detail in the next section.

|  | datajacket.org ($G_{dj}$) | D-Ocean ($G_{do}$) | D-Ocean (RS, $G_{rs}$) |
|---|---|---|---|
| # of nodes ($|N|$) | 798 | 2706 | 831 |
| # of links ($|E|$) | 16076 | 753443 | 72520 |
| Ave. degree ($\langle k \rangle$) | 40.3 | 556.8 | 174.5 |
| Density ($\rho$) | 0.051 | 0.205 | 0.210 |
| Ave. clustering coefficient ($\langle C \rangle$) | 0.702 | 0.891 | 0.893 |
| Assortativity ($r$) | 0.489 | 0.589 | 0.606 |
| Ave. path length ($\langle d \rangle$) | 3.26 | 2.28 | 2.31 |
| Diameter ($d^{max}$) | 10 | 7 | 5 |

Table 2: Characteristic values of the largest components of data networks

Furthermore, the average path lengths and diameters of $G_{do}$ are smaller than those of $G_{dj}$. $d_{rs}^{max}$ is half of $d_{dj}^{max}$ with the same number of nodes. This result suggests that many hub nodes appear and form very dense clusters with each other in $G_{do}$, which shrinks the network diameter. Figure 1(d) shows the degree distributions when the proportion of data with degree $k$ is $p(k)$. These rank-frequency plots are not in accordance with the typical power-law distributions. Both distributions exhibit a multi-scaling behavior, in which the low-degree nodes form gradual straight lines; however, the distribution declines abruptly when the degree increases. For $G_{do}$, the low-degree portion exhibits a horizontal straight line; however, the decline becomes abrupt when the degree reaches 272. In contrast, in the case of $G_{dj}$, a slow decay occurs at the beginning, and a rapid decay occurs at a degree of approximately 60, which is considerably rapid compared to that in the case of $G_{do}$. This finding indicates that some data groups exist in both the networks. In $G_{do}$, the distribution of the low-degree data represents a horizontal line and does not scale. In other words, the appearance ratio of the low-degree data is extremely small in $G_{do}$. The data start to scale from $k = 272$, and the hub nodes that acquire many links begin to appear. A large number of data is present in the network hubs, and the network has a globally and locally dense structure. The number of data points having the most links is 1537, which is connected to 56.8% of the data in the network. In other words, $G_{do}$ consists of a small number of low-degree data and a large number of high-degree data. $G_{dj}$, on the contrary, has a sparse structure in which a small number of hub data is connected to many data. Clusters are formed around the data that locally act as hubs, resulting in a locally dense structure. The highest degree is 172, which is connected to 21.6% of the data on the network. Most of the data have degrees between one and 100, and data $\leq 100$ degrees are extremely infrequent. In other words, $G_{dj}$ is composed of a large number of low-degree data and a small number of high-degree data.

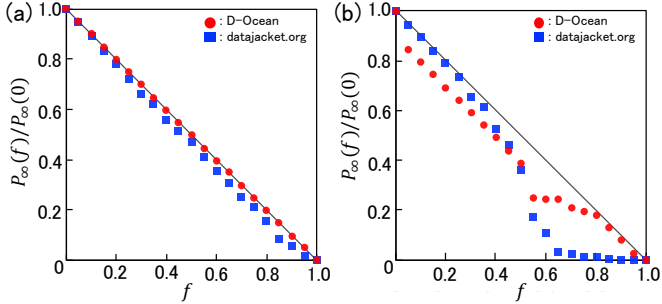

Figure 2: Robustness of the data networks

Figure 2 shows the robustness of each data network: Fig. 2 (a) shows the average of 10 independent processes subjected to random removal, and Fig. 2 (b) corresponds to the removal from high-degree nodes. Interestingly, despite the fact that $G_{dj}$ is globally sparse, it does not disintegrate at a certain threshold $f_c$, and finally disappears at $f = 1$, at which the behavior is almost similar to that of the global and local dense network of $G_{do}$. Therefore, almost all the nodes need to be removed to collapse the data networks. In other words, a data platform functions even if a certain number of data is lost at random. This is because the network hubs form a core group with a redundancy, and the assortative networks reduce the damage caused by the removal of the hub nodes [Vazquez and Moreno, 2003]. As shown in Fig. 2(b), $G_{dj}$ exhibits robustness even when data are removed from higher-degree nodes; this is equivalent to the characteristic of a scale-free network not being divided at the stage in which $f$ is small, owing to the presence of the redundancy. However, when 40% of the data are removed, major collapse begins to occur. In contrast, in the case of $G_{do}$, high-degree nodes decay sharply as soon as a 5% data loss occurs. However, after the first sharp decay, $G_{do}$ has only a monotonous decrease, and interestingly, the trend $P_\infty(f)/P_\infty(0)$ of $G_{dj}$ and $G_{do}$ reverses at approximately $f = 0.5$. The result shows that the hub nodes in $G_{do}$ function even with low-degree data, owing to the global dense feature. However, although $G_{dj}$ has a large number of low-degree data, none of the data points can be a hub; consequently, the redundancy is lost, and it decreases non-monotonically because of the global sparse feature.

## 4 Discussion

### 4.1 Data and Variable Structures

The variable-based structural analysis led to two main findings. First, despite the fact that the data are collected from widely different areas and different people, a power distribution appears in the data from both platforms. The experimental results show that the data platforms are composed not of data with many variables with high frequencies of appearance, but of data with a small number of variables with low frequencies. Regardless of the other aspects of data (storage formats, types, or size), the fact that the variety and number of variables in the data increases/decreases in proportion to a power law suggests that the same mechanism underlies both the small- and large-scale data, even in different data platforms.

Second, the data collection strategies are related to the data structures. The examination revealed that the maximum portion of datajacket.org data consists of data with few variables, and there is a diversity in the number of variables in the data compared with that of D-Ocean. Although the data of datajacket.org and D-Ocean have similar distributions at the macro-level, they are not completely equivalent and have different characteristics. First, the distributions of variables are different ($\gamma_{dj} > \gamma_{do}$), which shows that the datajacket.org data have few common variables, and most of the variables are unique to each type of data. The lack of communication at the time of data design may have affected these results. D-Ocean collects its data mainly from open data provided by different institutions. Some D-Ocean data, such as the data from governments, are designed uniformly, and many data have been published. The data holders of these shareable data can learn the design drawings of other data, as is the case with linked open data and open data whose schema is standardized. From the viewpoint of data design, the communication among data holders is relatively dense, which reduces the diversity of variables and the number of variables in each data. In contrast, the collection strategy of datajacket.org is different. It allows not only the metadata of shareable data, but also those of the sensitive data from individuals or private companies. These sensitive data are not to be made public or shared with others, and no design drawing of data (types of variables acquired as data) is shared. Therefore, most of the variables have low frequencies and are unique to the data, which increases the diversity of the variables and the number of variables in each data.

### 4.2 Network Structures

The network analysis resulted in two major findings. First, the network characteristics of the data platforms are more similar to a human relationship network than those of the natural world or engineering systems. In data networks, the links are provided between data pairs that have the same variables. Therefore, data that have similar variables tend to create clusters such as friend-of-friend interactions known to drive the dynamics of social networks. Owing to this mechanism, the networks of data platforms may behave similar to the human-related networks. The data network has the characteristics of an exponential distribution globally; however, a detailed investigation revealed that the distribution might pertain to a double-power-law, which also appears in the human communities [Ahn et al., 2007; Shin et al., 2015]. Considering that the network exhibits a positive high assortativity, only a few links exist between the low-degree and high-degree data groups, and each group of data forms a different community. The exponential distribution or the multi-scaling behavior result from the time limitations of people [Onnela et al., 2007] and the different types of users [Ahn et al., 2007]. Because there are no central figures who perform as super hubs for connecting with various other people, the networks have many small local groups with strong connections in the local communities. The limitation of human cognition for creating data may also represent characteristics that are similar to those of human networks. The data consist of variables recognized and used by humans. The data creators or collectors do not know all the data in the world, that is, data are not created according to a unified or widely shared design drawing. Therefore, even if the data pairs share some variables, not all of the data know one another's variables, resulting in a locally dense and highly assortative network.

Second, even if the data networks are similar each other, the data collection strategies influence the network characteristics and robustness. As discussed in the last subsection, datajacket.org collects heterogeneous data from diverse domains, and D-Ocean includes mainly open data, which results in a difference in the global density of the networks. The heterogeneous data strategy increases the diversity of the variables. However, although strong local connections exist, the entire network is sparse. Thus, the network seems fragile in terms of the robustness; however, the network of datajacket.org is redundant because it collects a variety of data from different organizations. In other words, even with the same data in the same area, the platform has several different data from different data holders, e.g., the supermarket provides apples from different farmers. Therefore, even if the hub data are lost, another similar data play the role of the hub, thereby maintaining the robustness of the network. In the open data strategy of D-Ocean, however, the platform loses the diversity of the variables but creates a network that is dense both globally and locally. The robustness of such a network is strong, and each node is a hub for the other. Therefore, even if most of the data is lost, the characteristics of the network are maintained, and the collapse of the network is smaller than that of datajacket.org. Regardless of the data platforms employing different strategies, the networks do not collapse even with a slight loss of data or removal of data users/holders, and the basic functions of the data platforms can be maintained. To summarize the strategies from the viewpoint of social application, from the viewpoint of robustness, a globally dense structure is preferable. However, from the viewpoint of marketability, it is desirable that the types of data and variables have diversity, and the open data strategy may lose the diversity from a market perspective, since only a few types of data are used. To optimize this trade-off is a future task.

## 5 Conclusion

The goal of this study was to deepen our understanding of the structural characteristics of the data on data exchange platforms and their relationships by using a variable-based structural analysis and network approach. By observing the characteristics of the data and variables, the data networks were noted to have a structure similar to the human relationship networks. Furthermore, no exceptional or typical data were found, and an extreme inequality existed in the data population. The fact that the variable frequency follows a long-tailed distribution regardless of the population size suggests that the same mechanism applies to both the small- and large-scale data on the platforms. To improve the understanding of the mechanism, in future work, a data growth model with different data collection strategies should be established based on simulations.

Data exchange platforms are new marketplaces that have only recently emerged. To enable the sustainable development of data exchange platforms in the digital economy, not only the robustness, but also the soundness and the resilience need to be discussed. Furthermore, several issues, such as the regulation, quality, and reliability of data must be addressed to understand the ecosystem of data platforms. We hope that our research will contribute to the development of data platforms in the future.

## Acknowledgments

This study is supported by MEXT Quantum Leap Flagship Program Grant Number JPMXS0118067246 and KAKENHI JP19H05577, Kyodo Printing Co., Ltd., and the Artificial Intelligence Research Promotion Foundation. We would like to thank D-Ocean, Inc. and avgidea, Inc. for sharing the data.